\documentclass[12pt]{article}
\usepackage{graphics}
\usepackage{amssymb,epsfig,amsmath,euscript,array,cite}

\newcounter{multieqs}

\newcommand{\be}{\begin{equation}}
\newcommand{\ee}{\end{equation}}
\newcommand{\eq}[1]{(\ref{#1})}

\newcommand{\bm}[1]{\mbox{\boldmath $#1$}}

\def\bd{\begin{document}}
\def\ed{\end{document}}
\def\nn{\nonumber}
\def\bea{\begin{eqnarray}}
\def\eea{\end{eqnarray}}
\let\bm=\bibitem
\let\la=\label

\def\npb#1#2#3{Nucl. Phys. {\bf{B#1}} #3 (#2)}
\def\plb#1#2#3{Phys. Lett. {\bf{#1B}} #3 (#2)}
\def\prl#1#2#3{Phys. Rev. Lett. {\bf{#1}} #3 (#2)}
\def\prd#1#2#3{Phys. Rev. {D \bf{#1}} #3 (#2)}
\def\cmp#1#2#3{Comm. Math. Phys. {\bf{#1}} #3 (#2)}
\def\cqg#1#2#3{Class. Quantum Grav. {\bf{#1}} #3 (#2)}
\def\nppsa#1#2#3{Nucl. Phys. B (Proc. Suppl.) {\bf{#1A}}#3 (#2)}
\def\ap#1#2#3{Ann. of Phys. {\bf{#1}} #3 (#2)}
\def\ijmp#1#2#3{Int. J. Mod. Phys. {\bf{A#1}} #3 (#2)}
\def\rmp#1#2#3{Rev. Mod. Phys. {\bf{#1}} #3 (#2)}
\def\mpla#1#2#3{Mod. Phys. Lett. {\bf A#1} #3 (#2)}
\def\jhep#1#2#3{J. High Energy Phys. {\bf #1} #3 (#2)}
\def\atmp#1#2#3{Adv. Theor. Math. Phys. {\bf #1} #3 (#2)}

\newcommand{\EQ}[1]{\begin{equation} #1 \end{equation}}
\newcommand{\AL}[1]{\begin{subequations}\begin{align} #1 \end{align}
\end{subequations}}
\newcommand{\SP}[1]{\begin{equation}\begin{split} #1 \end{split}\end{equation}}
\newcommand{\ALAT}[2]{\begin{subequations}\begin{alignat}{#1} #2
\end{alignat}\end{subequations}}
\def\beqa{\begin{eqnarray}}
\def\eeqa{\end{eqnarray}}
\def\beq{\begin{equation}}
\def\eeq{\end{equation}}

\def\N{{\cal N}}
\def\sst{\scriptscriptstyle}
\def\thetabar{\bar\theta}
\def\Tr{{\rm Tr}}
\def\one{\mbox{1 \kern-.59em {\rm l}}}

\def\a{\alpha}      \def\da{{\dot\alpha}}
\def\b{\beta}       \def\db{{\dot\beta}}
\def\c{\gamma}  \def\C{\Gamma}  \def\cdt{\dot\gamma}
\def\d{\delta}  \def\D{\Delta}  \def\ddt{\dot\delta}
\def\e{\epsilon}        \def\vare{\varepsilon}
\def\f{\phi}    \def\F{\Phi}    \def\vvf{\f}
\def\h{\eta}
\def\k{\kappa}
\def\l{\lambda} \def\L{\Lambda}
\def\m{\mu} \def\n{\nu}
\def\o{\omega}
\def\p{\pi} \def\P{\Pi}
\def\r{\rho}
\def\s{\sigma}  \def\S{\Sigma}
\def\t{\tau}
\def\th{\theta} \def\Th{\Theta} \def\vth{\vartheta}
\def\X{\Xeta}
\def\z{\zeta}

\def\cA{{\cal A}} \def\cB{{\cal B}} \def\cC{{\cal C}}
\def\cD{{\cal D}} \def\cE{{\cal E}} \def\cF{{\cal F}}
\def\cG{{\cal G}} \def\cH{{\cal H}} \def\cI{{\cal I}}
\def\cJ{{\cal J}} \def\cK{{\cal K}} \def\cL{{\cal L}}
\def\cM{{\cal M}} \def\cN{{\cal N}} \def\cO{{\cal O}}
\def\cP{{\cal P}} \def\cQ{{\cal Q}} \def\cR{{\cal R}}
\def\cS{{\cal S}} \def\cT{{\cal T}} \def\cU{{\cal U}}
\def\cV{{\cal V}} \def\cW{{\cal W}} \def\cX{{\cal X}}
\def\cy{{\cal y}} \def\cZ{{\cal Z}}

\def\ua{\underline{\alpha}}
\def\ub{\underline{\phantom{\alpha}}\!\!\!\beta}
\def\uc{\underline{\phantom{\alpha}}\!\!\!\gamma}
\def\um{\underline{\mu}}
\def\ud{\underline\delta}
\def\ue{\underline\epsilon}
\def\una{\underline a}\def\unA{\underline A}
\def\unb{\underline b}\def\unB{\underline B}
\def\unc{\underline c}\def\unC{\underline C}
\def\und{\underline d}\def\unD{\underline D}
\def\une{\underline e}\def\unE{\underline E}
\def\unf{\underline{\phantom{e}}\!\!\!\! f}\def\unF{\underline F}
\def\unm{\underline m}\def\unM{\underline M}
\def\unn{\underline n}\def\unN{\underline N}
\def\unp{\underline{\phantom{a}}\!\!\! p}\def\unP{\underline P}
\def\unq{\underline{\phantom{a}}\!\!\! q}
\def\unQ{\underline{\phantom{A}}\!\!\!\! Q}
\def\unH{\underline{H}}

\def\As {{A \hspace{-6.4pt} \slash}\;}
\def\bs {{b \hspace{-6.4pt} \slash}\;}
\def\Ds {{D \hspace{-6.4pt} \slash}\;}
\def\ds {{\del \hspace{-6.4pt} \slash}\;}
\def\ss {{\s \hspace{-6.4pt} \slash}\;}
\def\ks {{ k \hspace{-6.4pt} \slash}\;}
\def\ps {{p \hspace{-6.4pt} \slash}\;}
\def\pas {{{p_1} \hspace{-6.4pt} \slash}\;}
\def\pbs {{{p_2} \hspace{-6.4pt} \slash}\;}

\def\Fh{\hat{F}}
\def\Vh{\hat{V}}
\def\Xh{\hat{X}}
\def\ah{\hat{a}}
\def\xh{\hat{x}}
\def\yh{\hat{y}}
\def\ph{\hat{p}}
\def\xih{\hat{\xi}}

\def\psit{\tilde{\psi}}
\def\Psit{\tilde{\Psi}}
\def\tht{\tilde{\th}}

\def\At{\tilde{A}}
\def\Qt{\tilde{Q}}
\def\Rt{\tilde{R}}
\def\Nt{\tilde{N}}

\def\at{\tilde{a}}
\def\st{\tilde{s}}
\def\ft{\tilde{f}}
\def\pt{\tilde{p}}
\def\qt{\tilde{q}}
\def\vt{\tilde{v}}
\def\nt{\tilde{n}}

\def\delb{\bar{\partial}}
\def\bz{\bar{z}}
\def\bD{\bar{D}}
\def\bB{\bar{B}}

\def\bk{{\bf k}}
\def\bl{{\bf l}}
\def\bp{{\bf p}}
\def\bq{{\bf q}}
\def\br{{\bf r}}
\def\bx{{\bf x}}
\def\by{{\bf y}}
\def\bR{{\bf R}}
\def\bV{{\bf V}}

\def\d{\delta}\def\D{\Delta}\def\ddt{\dot\delta}

\def\pa{\partial} \def\del{\partial}
\def\xx{\times}
\def\uno{\mbox{1 \kern-.59em {\rm l}}}

\def\trp{^{\top}}
\def\inv{^{-1}}
\def\dag{{^{\dagger}}}
\def\pr{^{\prime}}

\def\rar{\rightarrow}
\def\lar{\leftarrow}
\def\lrar{\leftrightarrow}

\newcommand{\0}{\,\!}      
\def\one{1\!\!1\,\,}
\def\im{\imath}
\def\jm{\jmath}

\newcommand{\tr}{\mbox{tr}}
\newcommand{\slsh}[1]{/ \!\!\!\! #1}

\def\vac{|0\rangle}
\def\lvac{\langle 0|}

\def\hlf{\frac{1}{2}}
\def\ove#1{\frac{1}{#1}}

\def\Box{\square}
\def\ZZ{\mathbb{Z}}
\def\CC#1{({\bf #1})}
\def\bcomment#1{}
\def\bfhat#1{{\bf \hat{#1}}}
\def\VEV#1{\left\langle #1\right\rangle}

\newcommand{\ex}[1]{{\rm e}^{#1}} \def\ii{{\rm i}}

\newcommand{\lrbrk}[1]{\left(#1\right)}
\newcommand{\sfrac}[2]{{\textstyle\frac{#1}{#2}}}

\font\mybb=msbm10 at 12pt
\def\bb#1{\hbox{\mybb#1}}

\font\myBB=msbm10 at 18pt
\def\BB#1{\hbox{\myBB#1}}
\def\ap{\alpha^\prime}
\def\half{\frac{1}{2}}
\def\p{\partial}
\def\ba{\begin{eqnarray}}
\def\ea{\end{eqnarray}}

\newcommand*{\nind}{\noindent}

\begin{document}

\hfill{hep-th/0304191}

\vspace{20pt}

\begin{center}
{\Large \bf The String Light Cone in the pp-wave Background }
\vspace{30pt}

{\bf Chong-Sun Chu$^{a,}$\footnote{on leave of absence from
Department of Mathematics, University of Durham, UK},
Konstantinos Kyritsis$^{b}$
}

\vspace{15pt}
{\small \em
\begin{itemize}
\item[$^a$] Department of Physics, National Tsing Hua University, 
Hsinchu, Taiwan 300, R.O.C. \item[$^b$] Centre for
Particle Theory, Department of Mathematical Sciences  , 
University of Durham, Durham, DH1 3LE, UK
\end{itemize}
}

Email: {\sffamily \tt Chong-Sun.Chu@durham.ac.uk, 
Konstantinos.Kyritsis@durham.ac.uk}

\vspace{50pt} {\bf Abstract}

\end{center}

In this letter, we determine the particle and the 
string light cone in the pp-wave
background. The result is a deformed version of the flat one.
We point out the light cone exhibits an intriguing
periodicity in the light cone time direction $x^+$ 
with a period $\sim 1/ \mu$.
Our results also suggest that a quantum theory in the pp-wave background 
can be formulated consistently only if 
the background is periodic in the light cone time $x^+$. 

\vspace{0.5cm}
\setcounter{page}0
\newpage



\newpage


\section{Introduction}

Causality entered the realm of physics with the advent of special
relativity. The upper bound on the speed of any object (namely the
speed of light $c$) and the indefinite metric of the Minkowski
spacetime changed the idea of causality radically. One now would
speak, of events, points in spacetime. Given a present event, some
future events could be influenced by the propagation of a signal, some
others only by the propagation of a signal at the speed of light and
yet some others could not be influenced, simply because that would
require superluminal speeds.  And the idea of the light
cone emerged, defined simply as the hypersurface in spacetime that
divides causally related events from causally unrelated.
In the Minkowskian spacetime, the
light cone is defined by the proper distance $ds^2$ 
\be \label{flat-metric}
ds^2 = -2 dx^- dx^+ + dx^i dx^i
\ee 
and  divides the spacetime into regions where the proper distance
between events being time-like, light-like or space-like.

The marriage between quantum mechanics and special relativity proved to be
a non trivial one. After many attempts, this was accomplished with the
formulation of the quantum theory of fields.
The question of causality in a quantum field theory is formulated in terms
of the commutability of observables, or in terms of the
(anti-)commutability of its fundamental fields. The  light
cone in point particle quantum field theory, as defined by the vanishing of the
commutator of fundamental quantum fields,
is the same with the classical theory, 
as determined by special relativity.

String, as an extended object, is intrinsically different from a
particle. It is an interesting problem to study the notion of
causality in string theory and to see how is it different from the
particle case. This analysis has been performed for the case of  a
flat background \cite{Martinec:1993jq,Martinec:zv}, with the metric \eq{flat-metric}.
There, it was found
that the shape of the light cone is modified due to the extra,
internal oscillatory modes of the string. For example, two open
strings are causally unrelated (vanishing of the string
field commutator) if 
\be 2 \Delta
x^-_0 \Delta x^+ - \Delta {x^i}^2_0 - \sum_{l=1}^\infty | \Delta x^i_l
|^2 <0. \label{flatslc} 
\ee 
Here $x^\pm= (t \pm x) /\sqrt{2}$ and
$\Delta x = x - y$ for two strings $x(\sigma)$ and $y(\sigma)$. The
deviation from the usual understanding of light cone is obvious and is
due to the internal oscillatory modes that a string carries. This is indeed a
surprising result. Note that when the stringy contribution ($l\geq 1$
terms in \eq{flatslc}) is dropped, one
recovers the usual notion of light cone of a point particle, as
determined by the metric \eq{flat-metric}.
Effects of string interaction on the definition of
light cone has also been studied. Further
discussions of the result can be found  in \cite{Lowe1,Lowe2}.

Recently, string theory in pp-wave background have been studied with
immense interests, largely due to the remarkable proposal \cite{BMN}
which states that a sector of the SYM operators with large $R$-charge
is dual to the IIB string theory on a pp-wave background. The string
background consists of a  plane wave background, 
\be ds^2 = -2 dx^+ dx^- - \mu^2 \sum_{i=1}^8 (x^i)^2 dx^+
dx^+ + \sum_{i=1}^8 dx^i dx^i, \label{pp-wave metric}  
\ee  
together with a RR five-form. 
This background can be seen as a
deformation of the flat background, with a curvature controlled by
$\mu$. Remarkably, even in the presence of a RR 5-form flux, the
string theory is exactly solvable \cite{quan} in the light
cone gauge. It is therefore an interesting question to ask whether and
how the causal structure is modified in the pp-wave case. This is the
main motivation of this letter. We find that two strings in the pp-wave
background are causally unrelated if
\be \label{slc-0} 
\Delta x^-_0 - \frac{\mu}{2 \sin \left( \mu \Delta x^+
\right)} \sum_{i=1}^8 \sum_{l=0}^\infty 
\left[ \left( (x^i_l)^2 + (y^i_l)^2 \right) \cos \left( \mu
\Delta x^+ \right) - 2 x^i_l y^i_l
\right] < 0.
\ee

The plan of this letter is as follow. In section 2 we start  with a
particle propagating in the pp-wave background and use the commutator
algebra to extract the light cone. It is given by \eq{slc-0} by
striping out all the stringy modes ($l\geq 1$) contribution. 
The particle light cone is exactly the same as
the one determined by the metric \eq{pp-wave metric}. 
After this warm up exercise, we
construct in section 3 the light cone string field in the pp-wave
background. We then compute the string field commutator and use it to
extract the string light cone. Discussions and further comments are
found in section 4.

\section{Particle light cone in the  pp-wave background}

Consider a free relativistic real scalar particle moving in the metric
\eq{pp-wave metric}. The light cone Hamiltonian is given by
\be \label{particle Hamiltonian} 
H = \frac{1}{2p^+}  
\left( p^{i \, 2} + m^2  + \o^2 x^{i \, 2} \right), \quad \o:= \mu p^+   .
\ee 
Notice that it consists of two parts. The first is
the usual piece for a free massive relativistic particle and the other
is an oscillator's potential.

The field for the particle has to obey the Schr\"{o}ndinger equation
\be 
 i \frac{\p \phi}{\p x^+}=H \phi,   
\ee   
In the coordinate
representation, the equation of motion reads  
\be  \label{eom-p}
i \frac{\p\phi}{\p x^+} = 
\frac{1}{2p^+} \left( -\frac{\p^2}{{\p x^i}^2} + m^2 + \o^2 x^{i \, 2} \right) 
\phi := \frac{H_0}{2p^+} \phi .  
\ee  
$H_0$ is the Hamiltonian for an 
oscillator with mass 1/2 and a frequency $2\o$. The light cone energy is  
\be
 H = \sum_i n^i \mu  + \frac{m^2}{2 p^+}  
\ee 
where we have dropped the zero point energy. Note that the
physical mass $m$ of the particle appears as a constant energy. We
will see below that the light cone does not depend on $m$.

The equation \eq{eom-p} can be solved (separate variables) and at the
end, the solution one finds for the field is 
\ba 
\phi (x^+, x^-, x^i)&
= & \int_0^\infty \frac{dp^+}{\sqrt{2\pi p^+}}  \sum_{\{n^i\}} a\left(
p^+, \{n^i\} \right) e^{-i(x^+ p^- + x^- p^+)} \cdot \nn \\ & & \cdot
\prod_{i=1}^8 H_{n^i}\left( \sqrt{\o} x^i\right)  \exp\left[ - \half
\o (x^i)^2  \right] \sqrt{\frac{\sqrt{\o/\pi}}{2^{n^i} \left( n^i \right)!}} 
\quad + h.c. 
\ea 
Demanding the equal
time commutator  
\be \left[ \phi(x^+,x^-,x^i),\phi(x^+,y^-,y^i)
\right] = \delta(x^- - y^-) \prod_{i=1}^8 \delta(x^i - y^i) ,  
\ee
implies the commutation relation for the creation-annihilation
operators  
\be \left[ a\left(p^+,\{n^i\}\right), a^\dag
\left(q^+,\{m^j\} \right) \right]  = p^+ \delta(p^+ - q^+)
\delta_{\{n^i\},\{m^j\}}  
\ee 
and the other being zero.

We are now in position to find the propagator for the particle. It is
easy to obtain  
\be  
\left[ \phi(x^+,x^-,x^i),\phi(y^+,y^-,y^i)\right]
= I_1 -I_2  
\ee  
where $I_1 = I_1(\D x^+, \D x^-, x^i,y^i)$ is given by  
\be  
I_1 =\int_0^\infty \frac{dp^+}{2\pi} e^{-i \Delta x^- p^+ - i
\Delta x^+ \frac{m^2}{2p^+}} \prod_{i=1}^8 \sum_{n^i} e^{-i \mu \D x^+
n^i} H_{n^i}(\sqrt{\o} x^i) H_{n^i}(\sqrt{\o} y^i) e^{-\half \o \left(
(x^i)^2 + (y^i)^2)  \right) } \frac{\sqrt{\o/\pi}}{2^{n^i}  \left( n^i
! \right)}  \nn  
\ee  
and $I_2 = I_1^*$. Here $\D x^+ := x^+ - y^+$,
$\Delta x^- := x^- - y^-$. With the propagator in our hands,  we are
in a position to find the light cone for the point particle.  Instead of
considering an oscillating phase factor,  the integral can be computed
by an analytical continuation of $p^+$. Without loss of generality, we
assume $\D x^+ >0$  and we perform an analytic continuation  
\be p^+
\longrightarrow i p^+ \quad \mbox{for $I_1$}.  
\ee 
Then  
\bea
\label{p-I1} 
I_1 &=& \int_0^\infty \frac{dp^+}{2\pi} e^{\Delta x^- p^+
- \Delta x^+ \frac{m^2}{2p^+} } 
\prod_{i=1}^8 \frac{\sqrt{\mu p^+ /\pi}}{\sqrt{1 - e^{-2i\mu \Delta x^+ }}} 
\cdot \nn\\ && 
\cdot \exp
	\left\{ \frac{\mu p^+ }{2 \sin \left(\mu \Delta x^+ \right)}
	\left[ 2 x^i y^i - \left( {x^i}^2 + {y^i}^2 \right) \cos
	\left( \mu\Delta x^+ \right) \right] \right\} .  
\eea
Performing an analytic continuation  
\be 
p^+ \longrightarrow - i p^+ \quad \mbox{for $I_2$},  
\ee  
one obtains the same identical
expression for $I_2$. The integral converges in the $p^+ \rightarrow
\infty$ limit, and hence the commutator \eq{comm} vanishes, if the
exponent factor is negative, i.e. 
\be 
\Delta x^- - \frac{\mu}{2
\sin(\mu \D x^+)} \sum_{i=1}^8\left[ \left( (x^i)^2 + (y^i)^2 \right)
\cos(\mu\Delta x^+) - 2 x^i y^i \right] < 0 . 
\label{point particle lc}
\ee
This is the equation for the point particle light cone in a pp-wave
spacetime. 

A couple of remarks are in order. First we note  
that the light cone \eq{point particle lc} is independent of the mass
$m$ of the 
particle,  as one can expect on physical grounds. 
Secondly, we note that the particle light cone \eq{point particle lc}
is exactly the same as the light cone determined by the metric
\eq{pp-wave metric} \footnote{We thank Harald Dorn, Boris Pioline and
  Simon Ross  
for pointing out a mistaken statement 
made in the first version of this letter.}. See \cite{simon} for interesting 
discussions of the properties of light cone and causality 
as pertained to the pp-wave metric. 
Finally, we note also that the
light cone is periodic in $\D x^+ \sim \D x^+ + 2 \pi/\mu$. We will
see that the same 
intriguing behaviour persists in the string case.

\section{String light cone in the pp-wave background}

The construction of a light cone string field theory in this new
background follows the same steps with the construction in the flat
background, see for example \cite{Kaku:zz}.

Consider a bosonic open string moving in a pp-wave background,  with
metric given by (\ref{pp-wave metric}). The $i$ index enumerates the
transverse coordinates, $i=1,\ldots, d-2$. The generalization to
closed string and to the supersymmetric case is straightforward. By
taking a light cone gauge 
\be \label{lc-cond} 
x^+ = p^+ \tau, 
\ee 
the theory is exactly solvable,  much like the case of  a string in a
flat, Minkowski, background. For a more detailed discussion, see
\cite{quan,CH}. In the light cone gauge, the string
coordinates $x^-$ and the transverse ones $x^i$ are given by 
\bea 
x^-(\s) = x^-_0 + \int^\s d \tilde{\s} \;  {x^i}' P^i(\tilde{\s}) \\
x^i(\sigma) =  x^i_0 + \sqrt{2} \sum_{l=1}^\infty x^i_l \cos (l\sigma).
\label{mode open}
\eea
The light cone string field $\Phi$ depends on  $x^+, x_0^-, x^i(\s)$,
i.e. $\Phi = \Phi[x^+ , x_0^-, x^i(\s)]$ and satisfies the
Schr\"{o}dinger  equation 
\be 
\label{sch-eq1} i \frac{\p \Phi}{\p x^+}
= H \Phi[x^+ , x_0^-, {x^i}(\s)], 
\ee 
where $H$ is the light cone
Hamiltonian (taking $2\a'=1$) 
\be 
H = \frac{\pi }{2 p^+} \int_0^\pi
d\sigma \left[ -\frac{\p^2}{\p {x^i}(\sigma)^2} + \frac{1}{\pi^2}
\left( \frac{\del x^i(\sigma)}{\del \s}\right)^2 + \frac{m^2}{\pi^2} (x^i(\sigma))^2
\right] \quad \mbox{and} \quad m := \mu p^+. 
\label{Hamiltonian} 
\ee

It is straight forward to rewrite $H$ is terms of modes and
\eq{sch-eq1} then takes the form 
\be 
i \frac{\p \Phi}{\p x^+} =
\frac{1}{2p^+} \sum_{l=0}^\infty H_l \ \Phi(x^+, x^-, \{x^i_l\}) ,
\label{sft eom modes}
\ee 
where 
\be H_l := -\frac{\p^2}{\p {x^i_l}^2} +
\o_l^2 (x^i_l)^2, \quad \o_l :=\sqrt{l^2 + (\mu p^+)^2
}. 
\label{Hamiltonian modes} 
\ee 
Clearly, (\ref{Hamiltonian modes}) is
a SHO Hamiltonian with  mass 1/2 and frequency $2 \o_l$.
Eq. \eq{sft eom modes} can be solved easily and the final answer for
the string field is 
\be 
\Phi = \int \frac{dp^+}{\sqrt{2\pi p^+}}
\sum_{\{n^i_l\}} A\left( p^+, \{n^i_l\} \right) e^{-i \left( x^+ p^- +
x^- p^+ \right)} \prod_{l=0}^\infty \varphi^l_{\{n^i_l\}} ( x^i_l ) +
h.c. 
\ee 
where 
\be 
\varphi^l_{\{n^i_l\}} ( x^i_l )= \prod_{i=1}^{d-2}
H_{\{n^i_l\}} \left( \sqrt{\omega_l} x^i_l \right) e^{-\omega_l
({x^i_l})^2 / 2} \sqrt{ \frac{ \sqrt{\omega_l/\pi}}{2^{ n^i_l} (n^i_l!)}} .
\ee 
The light cone energy is 
\be 
p^- = \frac{1}{p^+}
 \sum_{i=1}^{d-2} \sum_{l=0}^\infty n^i_l \omega_l, 
\label{mass-shell}
\ee 
where we have ignored the zero point energy, which is
irrelevant for our analysis. The conjugate momentum field is given by,
see \cite{Kaku:zz}, 
\be 
\Pi[x(\sigma)] = \frac{ \delta \mathcal{L} }{
\delta \left(\p\Phi / \p x^+\right) } 
\ee and in the present case it
is \be 
\Pi = i \Phi. 
\ee

The Equal Time Commutation Relations for the string field are 
\be
\left[ \Phi\left(x^+,x^-_0,\vec{x_l}\right),
\Phi\left(x^+,y^-_0,\vec{y_l}\right)\right] = \delta(x^-_0 - y^-_0)
\prod_{i=1}^{d-2} \delta\left[ x^i(\sigma) -
y^i(\sigma)\right]. 
\label{ECTR} 
\ee 
which translate to the
commutation relation 
\be 
\left[ A\left(p^+, \{n^i_l\} \right), A^\dag
\left( q^+, \{m^j_k\} \right) \right] = p^+ \delta(p^+ - q^+) \delta_{
\{n^i_l\},\{m^j_k\}} 
\label{cm relations} 
\ee for the string
creation-annihilation operators. The calculation of the general
(non-equal time) commutator of the string field is not hard to perform
and one finds at the end that 
\be 
\label{comm} \left[
\Phi(x^+,x^-_0,x^i_l),  \Phi(y^+,y^-_0,y^i_l)\right] = I_1 - I_2 ,
\ee
where $I_1=I_1(\D x^+, \D x^-_0,x^i_l, y^i_l)$ is given by 
\bea
\label{propagator} 
I_1 &:=& \int_0^\infty \frac{dp^+}{2\pi} e^{-i
\Delta x^-_0 p^+} \prod_{i}\prod_{l=0}^{\infty} \frac{\sqrt{\o_l/\pi}}
{\sqrt{1- e^{-i \frac{2\Delta
x^+ \omega_l}{p^+}}}} \cdot \nn\\ 
&& \cdot \exp \left\{
\frac{\omega_l}{2i \sin \left( \frac{\Delta x^+ \omega_l}{ p^+}
\right)} \left[ 2 x^i_l y^i_l - \left( (x^i_l)^2 + (y^i_l)^2 \right)
\cos \left(\frac{\Delta x^+ \omega_l}{p^+}\right) \right] \right\}
\eea 
and $I_2(\D x^+, \D x^-_0,x^i_l, y^i_l) = I_1(-\D x^+, -\D
x^-_0,x^i_l, y^i_l) = [I_1(\D x^+, \D x^-_0,x^i_l, y^i_l) ]^*$.  
Notice that as expected, the propagator
is a product of SHO propagators.

To compute $I_1$ and $I_2$, it is convenient to analytic continuate
$p^+$ to the complex  plane. Without loss of generality, we assume $\D
x^+ >0$ and perform an analytic continuation 
\be 
p^+ \longrightarrow i
p^+ \quad \mbox{for $I_1$}. 
\ee 
Then 
\bea \label{I1} I_1 &=&
\int_0^\infty \frac{dp^+}{2\pi} e^{\Delta x^-_0 p^+} \prod_{i}\prod_{l=0}^{\infty}
\frac{\sqrt{\o'_l/\pi}}{\sqrt{1 - e^{ - \frac{2\Delta x^+ \omega_l'}{p^+}}}} \cdot
\nn\\ && \cdot \exp \left\{ \frac{\omega_l'}{2 \sinh \left(
\frac{\Delta x^+ \omega_l'}{p^+} \right)} \left[ 2 x^i_l y^i_l -
\left( (x^i_l)^2 + (y^i_l)^2 \right) \cosh \left( \frac{\Delta x^+
\omega_l'}{p^+} \right) \right] \right\} , 
\eea 
where 
\be \omega_l' =
\sqrt{l^2 - (\mu p^+)^2} . 
\ee 
Performing an analytic continuation 
\be
p^+ \longrightarrow - i p^+ \quad \mbox{for $I_2$}, 
\ee 
one obtains the same identical expression for $I_2$.

Therefore the commutator \eq{comm} vanishes provided that the integral
\eq{I1} converges. To check when this is the case,  note that the
integral converges for $p^+ \rightarrow 0$. However problems may
appear in the regime $p^+ \rightarrow \infty$. In this regime,
$\omega_l'$ becomes imaginary 
\be 
\omega_l' = i \sqrt{(\mu p^+)^2 - l^2} \longrightarrow i \mu p^+
\ee Therefore \eq{I1} becomes 
\ba  
I_1 
& = & \int_0^\infty
\frac{dp^+}{2\pi} e^{\Delta x^-_0 p^+}
\prod_{i}\prod_{l=0}^{\infty}
\frac{\sqrt{\mu p^+/\pi}}{\sqrt{1 - e^{-2 i  \mu \Delta x^+ }}} 
\exp \left\{ \frac{ \mu p^+}{2 \sin \left( \mu \Delta x^+  \right)} 
\left[ 2 x^i_l y^i_l - \left( (x^i_l)^2 + (y^i_l)^2 \right) 
\cos \left(\mu \Delta x^+ \right) \right] \right\}
\nn \\ 
&\sim & \int^\infty dp^+ \exp  \left\{ p^+ \left[\Delta x^-_0 -
\frac{\mu}{2 \sin \left(\mu \Delta x^+ \right)} \sum_{i,l} \left[
\left( (x^i_l)^2 + (y^i_l)^2 \right) \cos \left( \mu \Delta x^+
\right) - 2 x^i_l y^i_l \right] \right]\right\} 
\ea 
The integral
converges in the $p^+ \rightarrow \infty$ limit, and hence the
commutator \eq{comm} vanishes, if the exponent factor is negative,
i.e. if 
\be 
\Delta x^-_0 - \frac{\mu}{2 \sin \left( \mu \Delta x^+
\right)} \sum_{i}\sum_{l=0}^\infty \left[ \left( (x^i_l)^2 + (y^i_l)^2 \right) \cos
\left( \mu \Delta x^+ \right) - 2 x^i_l y^i_l \right] < 0. \label{slc}
\ee 
Equation (\ref{slc}) determines the string light cone in the pp-wave
background and is the main result of this letter.

 Notice that the string light cone \eq{slc} is determined not only by
the zero modes, but also by the internal oscillating modes of the
string. This is something that we should expect since the same phenomenon
appears in the flat background case. Keeping only the zero mode
contribution, \eq{slc} reduces to 
\eq{point particle lc} the light cone for the point particle in the same
background.

\section{Conclusions}

We make a couple of remarks.

{\bf 1.} 
Unlike the flat case, the string light cone in the pp-wave
case is not a function of $x^i_l - y^i_l$ any more and hence
translational invariance is lost. But this should not come as a
surprise. The pp--wave metric is not translational invariant itself,
so the same should apply to the light cone as well. This is consistent
with what we found.

{\bf 2.} We note that in the limit $\mu \rightarrow 0$, we recover the
flat background expression for the light cone, equation
(\ref{flatslc}). This is expected since the metric (\ref{pp-wave
metric}) in the $\mu \rightarrow 0$ limit reduces to the Minkowski
metric. For a nonzero $\mu$ and in a small region of $x^+$ such that
\be 
|\Delta x^+| \ll \frac{1}{\mu} , 
\ee we are probing the part of
the spacetime that is  very close to the original string itself. In
this limit, spacetime curvature can be ignored. Physically one expects
that the derivation from the flat case to be small and the light
cone for the flat case \eq{flatslc} to be recovered. This is indeed
the case. The light cone \eq{slc} is a continuous function of $\mu$
and matches up with the flat case.


{\bf 3.}
In this letter, we considered a free string. It would be interesting to
include interactions (tree level and loops) and to study how the string
light cone gets modified, see \cite{Lowe1,Lowe2} for a tree level
computation in the flat case.  String interaction in pp-wave
background has been considered in \cite{vsp}, and reconsidered in
\cite{sft}
where the $Z_2$ symmetry of the background is taken into account
properly. Field theory analysis of the $Z_2$ symmetry was recently
performed in \cite{z2}. Our result agrees with the $Z_2$ symmetry of
the background. It will be interesting to examine this issue by including
the $Z_2$ invariant interaction.

{\bf 4.} Finally we note that the light cone \eq{slc} is periodic with 
$\D x^+ \sim \D x^+ + 2\pi/\mu$. The follows from a periodicity of 
the light cone time
\be \label{periodic}
 x^+ \sim x^+ + 2\pi/\mu.
\ee
However we did not impose any periodic
boundary condition on the metric
\footnote{Note that the if one think of the pp-wave metric 
\eq{pp-wave metric} as arised from the Pernose limit \cite{blau}, 
then by definition
$x^+$ is automatically periodic with periodicity $2\pi/\mu$. 
CSC thanks Rodolfo Russo for bringing this very interesting 
observation to his attention.} to start with. 
Our result suggests that to have a consistent quantum theory on the 
pp-wave background, the background should have this periodicity \eq{periodic}.
See \cite{DLCQ} for related works on discrete light cone quantization (DLCQ) of
string in pp-wave background.
Also we do not understand yet the physical significance of the 
periodicity of the causal structure of spacetime 
\footnote{Simon Ross suggests to us that, apart from the periodicity
interpretation,  since \eq{point particle lc}
and \eq{slc} is singular at $\D x^+ = \pi/\mu$, one may argue that
there is a relationship between pasts and futures shifted by $\pi/\mu$.
See the first paper in \cite{simon} for more detailed discussions.}.
We expect this to lead to extremely interesting physics. 
We leave these issues 
for further studies.

\section*{Acknowledgements}

CSC would like to thank Bin Chen, Harald Dorn, 
Valya Khoze, Feng-Li Lin, Boris Pioline, Simon Ross, Rodolfo Russo 
and Gabriele Travaglini 
for interesting discussions and useful comments. KK would
like to acknowledge George Georgiou and Apostolos Dimitriadis for interesting
discussions and useful comments. We acknowledge grants from EPSRC,
Nuffield foundation of UK, NSC of Taiwan and University of Durham.

\end{document}